\documentclass[12pt]{article}

\usepackage{amsmath,epsfig}

\begin{document}
\hfill IU-MSTP/76

\hfill Dec, 2008

\vskip 2cm
\begin{center}
{\Large\bf Magnetic translation symmetry 
on the lattice}
\end{center}

\vspace*{1cm}
\def\thefootnote{\fnsymbol{footnote}}
\begin{center}{\sc Ken-ichi Sekiguchi${}^1$, 
Tomohiro Okamoto${}^1$  and Takanori Fujiwara${}^2$}
\end{center}
\vspace*{0.2cm}
\begin{center}
{\it ${}^1$ Graduate School of Science and Engineering, Ibaraki University,
Mito 310-8512, Japan \\
${}^2$ Department of Physics, Ibaraki University,
Mito 310-8512, Japan}
\end{center}

\vfill
\begin{center}
{\large\sc Abstract}
\end{center}
\noindent
Magnetic translation symmetry on a finite periodic square 
lattice is investigated for an arbitrary uniform magnetic 
field in arbitrary dimensions. It can be used to classify 
eigenvectors of the Hamiltonian. The system can be 
converted to another system of half or lower 
dimensions. A higher dimensional generalization of Harper 
equation is obtained for tight-binding systems. 
\vfil
\newpage
\pagestyle{plain}
\section{Introduction}
\label{sec:intro}
\setcounter{equation}{0}

Systems of charged particles interacting with uniform 
magnetic fields have served as testing grounds of various 
ideas in quantum field theory and condensed matter physics. 
Uniform magnetic fields are simple and in many cases exact 
solutions are available while keeping some essential 
aspects of the interactions with the gauge fields. 
A remarkable feature of these system is the appearance 
of magnetic translation symmetry \cite{Zak}. 

In the presence 
of magnetic fields the gauge potential depends on the 
coordinates. The change in the gauge potential 
due to a translation can be compensated by a suitable 
gauge transformation. Arbitrary translations become a 
symmetry of the Hamiltonian in infinite euclidean 
spaces and each eigenstate is infinitely degenerated. 
In finite periodic spaces or on compact tori the boundary 
conditions for the wave functions become twisted. 
Because of this only a discrete finite subgroup of 
the whole translations survives as magnetic translation 
symmetry. The degeneracy of states is a finte constant 
depending on the magnetic field. 

On finite periodic lattices exact solutions are not 
available and translation symmetries are 
rather restricted. The spectra are far richer than 
those of the continuum \cite{hof}. Nevertheless, it is possible 
to define magnetic translations, which enable us to 
qualitatively understand the richness of the spectra. 
In this paper we investigate magnetic translation 
symmetry on the lattice in full detail for arbitrary 
uniform magnetic fields in arbitrary higher dimensions. 
The key idea is to introduce oblique lattice 
where the magnetic field takes a block-diagonal form 
composed of $2\times2$ anti-symmetric matrices. 
According to the rank of magnetic field, the 
generators of magnetic translations can be classified 
into pairs satisfying two dimensional magnetic 
translation algebra and unpaired commuting generators. 
This enables us to identify maximal commuting 
subset of the generators. Representation theoretic 
approach \cite{Sakamoto-Tanimura} can be used to 
reduce Hamiltonians into 
block-diagonal forms without relying on its 
concrete form. The reduced Hamiltonian describes 
a system of half or lower than half dimensions. 
This phenomenon has been observed for systems 
of electrons in periodic potentials 
\cite{Schellnhuber-Obermair}.
Magnetic translation symmetry has been investigated 
in four dimensions in the context of chiral lattice 
fermions \cite{GG-AHNR}. Reduction from two to one 
dimension and the spectra of hermitian Wilson-Dirac 
operator have also been discussed in Ref. 
\cite{fujiwara}. 

This paper is organized as follows. In the next 
section we show that periodic boundary conditions 
on finite lattice can be converted to twisted 
periodicity on infinite lattice. In Sect. 
\ref{sec:lmtr} we examine magnetic translations 
on the lattice. Constraints on the translation 
vectors are solved. To illustrate the approach 
we argue magnetic translation symmetry in two 
dimensions. This is done in Sect. 
\ref{sec:mtr2dl}. We introduce oblique lattice 
coordinates in Sect. \ref{sec:olc} and give a 
generalization to arbitrary dimensions. 
Sect. \ref{sec:sd} is devoted to summary and 
discussion.

\section{Twisted boundary conditions}
\label{sec:tbc}
\setcounter{equation}{0}

Let us consider a system of a charged particle 
interacting with a uniform magnetic field 
of an abelian gauge theory on a $d$ dimensional 
periodic lattice of a size $L^d$, where $L$ is 
a positive integer. We take the lattice constant 
$a=1$. Any uniform magnetic field 
$F_{\mu\nu}=-F_{\nu\mu}$ ($\mu,\nu=1,\cdots,d$)
on the lattice can be written as 
\begin{eqnarray}
  \label{eq:m2F}
  F_{\mu\nu}=\frac{2\pi m_{\mu\nu}}{L^2}, 
\end{eqnarray}
where $m_{\mu\nu}$ are integers. These classify 
topological sectors of the lattice gauge fields 
\cite{lus, fsw}. 

The link variables $U_\mu(x)$ giving rise 
to the uniform magnetic field can be chosen as 
\cite{lus} 
\begin{eqnarray}
  \label{eq:lv}
    U_\mu(x)=\exp\left[-\frac{2\pi i}{L}
      \delta_{\bar x_\mu,L-1}
      \sum_{\nu>\mu}m_{\mu\nu}\bar x_\nu
      -\frac{2\pi i}{L^2}\sum_{\nu<\mu}m_{\mu\nu}
      \bar x_\nu+ib_\mu\right],
\end{eqnarray}
where $0\leq\bar x_\mu<L$ ($\mu=1,\cdots,d$) 
stand for periodic lattice 
coordinates satisfying $\overline{x_\mu+L}
=\bar x_\mu$ and $b_\mu$ are real constants. 
These satisfy periodic boundary conditions
\begin{eqnarray}
  \label{eq:lvpbc}
  U_\mu(x+L\hat\nu)&=&U_\mu(x), 
\end{eqnarray}
where $\hat\nu$ is the unit vector along the 
$\nu$-th lattice axis. 
The magnetic fields (\ref{eq:m2F}) are related 
to the plaquette variables by 
\begin{eqnarray}
  \label{eq:plq}
  U_\mu(x)U_\nu(x+\hat\mu)
  U^\dagger_\mu(x+\hat\nu)
  U^\dagger_\nu(x)=e^{iF_{\mu\nu}}.
\end{eqnarray}

We denote the Hamiltonian or the lattice Dirac 
operator of the system by $H$ and consider 
the eigenvalue problem 
\begin{eqnarray}
  \label{eq:eveq}
  H\psi(x)&=&\lambda \psi(x).
\end{eqnarray}
Concrete form of $H$ is not necessary 
to argue magnetic translation symmetries. We only 
assume that $H$ only depends on the lattice 
coordinates through the link variables. 
The wave function $\psi(x)$ is subject to the 
the periodic boundary conditions
\begin{eqnarray}
  \label{eq:pbc}
  \psi(x+L\hat\mu)&=&\psi(x). 
\end{eqnarray}

In lattice gauge theory it is customary to adopt 
periodic boundary conditions. The lattice  is 
supposed to be a regularization of the infinite 
continuum and the boundary conditions become 
irrelevant in the infinite volume limit. 
In the present case the lattice should be 
understood as a regularization of a finite 
volume flat torus and the periodic boundary 
condition is considered to be legitimate. 
In the continuum, however, neither parallel 
transporters, the continuum analog of lattice 
link variables, nor wave functions can be 
periodic on tori in the presence of a net 
magnetic flux. 

This apparent mismatch between the continuum 
and the lattice can be resolved by noting 
that the periodic link variables  
(\ref{eq:lv}) become singular in the classical 
continuum limit $a\rightarrow0$. The singularities can be 
removed by a gauge transformation 
\begin{eqnarray}
  \label{eq:lsgtr}
  \Lambda_0(x)&=&\exp\left[-2\pi i
    \sum_{\mu<\nu}m_{\mu\nu}\left[\frac{x_\mu}{L}\right]
    \frac{x_\nu}{L}\right],
\end{eqnarray}
where $[c]$ stands for the integer part of $c$, i.e., 
$[c]=n$ if $n\leq c<n+1$ for some integer $n$. 
The $\Lambda_0(x)$ removes the troublesome parts 
of (\ref{eq:lv}) and yields 
\begin{eqnarray}
  \label{eq:nplv}
  U^{\rm a}_\mu(x)&=&\Lambda_0(x)U_\mu(x)
  \Lambda^\dagger_0(x+\hat\mu) \nonumber \\
  &=&\exp\left[-\frac{2\pi i}{L^2}\sum_{\nu<\mu}
    m_{\mu\nu}x_\nu+ib_\mu\right] \nonumber \\
  &=&\exp\left[-i\sum_{\nu<\mu}
    F_{\mu\nu}x_\nu+ib_\mu\right].
\end{eqnarray} 
This link variable has a smooth classical continuum 
limit and (\ref{eq:plq}) becomes apparent in this gauge. The cost 
we must pay is the simple periodicity of the link variable. 
The gauge transformation (\ref{eq:lsgtr}) is 
not periodic under the shift $x\rightarrow x+L\hat\mu$ 
but satisfies 
\begin{eqnarray}
  \label{eq:Lmu}
  \Lambda_0(x+L\hat\mu)=\Lambda^{\rm a}_\mu(x)\Lambda_0(x). \qquad
  \left(\Lambda^{\rm a}_\mu(x)=\exp\left[-iL\sum_{\nu>\mu}
    F_{\mu\nu}x_\nu\right]\right) 
\end{eqnarray}
It changes the periodic boundary conditions to 
twisted ones. The gauge transformed wave functions 
\begin{eqnarray}
  \label{eq:nnpwf}
    \psi^{\rm a}(x)&=&\Lambda_0(x)\psi(x) 
\end{eqnarray}
are subject to 
\begin{eqnarray}
  \label{eq:tbcp}
  \psi^{\rm a}(x+L\hat\mu)&=&\Lambda^{\rm a}_\mu(x)\psi^{\rm a}(x). 
\end{eqnarray}

The link variable (\ref{eq:nplv}) corresponds to  
axial gauge in the continuum. We can also work in the 
symmetric gauge by carrying out a further gauge 
transformation by
\begin{eqnarray}
  \label{eq:omega}
  \Lambda^{\rm sa}(x)&=&\exp\left[
    \frac{i}{2}\sum_{\mu<\nu}F_{\mu\nu}
    x_\mu x_\nu\right].
\end{eqnarray}
The link variable in the symmetric gauge is given by 
\begin{eqnarray}
  \label{eq:lvsg}
  U^{\rm s}_\mu(x)=\exp\left[-\frac{i}{2}\sum_{\nu}
    F_{\mu\nu}x_\nu+ib_\mu\right]. 
\end{eqnarray}
In this gauge the link variables and the wave functions 
satisfy the boundary conditions 
\begin{eqnarray}
  \label{eq:tbcsg}
  U^{\rm s}_\mu(x+L\hat\nu)
  =\Lambda^{\rm s}_\nu(x)
  U^{\rm s}_\mu(x)\Lambda^{{\rm s}\dagger}_\nu(x+\hat\mu), \qquad
  \psi^{\rm s}(x+L\hat\mu)&=&\Lambda^{\rm s}_\mu(x)\psi^{\rm s}(x), 
\end{eqnarray}
where $\Lambda^{\rm s}_\mu(x)$ is defined by 
\begin{eqnarray}
  \label{eq:Lsmu}
  \Lambda^{\rm s}_\mu(x)=
    \exp\left[-\frac{iL}{2}\sum_\nu 
      F_{\mu\nu}x_\nu\right].
\end{eqnarray}
This gauge is suitable for analyzing magnetic 
translation symmetry 
in the next section. 

The constant phases $b_\mu$ in the link variables 
do not affect the magnetic field $F_{\mu\nu}$. 
In the continuum we can remove them by a suitable 
choice of the coordinate origin if 
$\det F_{\mu\nu}\ne0$. The spectrum of the 
Hamiltonian does not depend on them. This is not the case 
for $\det F_{\mu\nu}=0$ as is observed in odd dimensions. 
In lattice gauge theory we can also remove $b_\mu$ from 
the link variables by a further twisting of the boundary 
conditions. The eigenvalues depend on $b_\mu$ even 
if $\det F_{\mu\nu}\ne0$ is satisfied. In this section we 
have not carried out this twisting, leaving the constant 
term intact.

\section{Magnetic translation symmetry on the lattice}
\label{sec:lmtr}
\setcounter{equation}{0}

In the continuum the Hamiltonian of a charged 
particle in a uniform magnetic field is not invariant 
under arbitrary translations. 
This is due to the coordinate dependence of the gauge 
potential. A discrete subgroup of the translations 
known as the magnetic translation \cite{Zak}, however, survives. 
It is a suitable combination of translations and 
gauge transformations. We expect that a similar 
situation also occurs on the lattice. In this section 
we pursue the conditions for the magnetic translations
in lattice gauge theory. 

We work in the symmetric gauge (\ref{eq:lvsg}) and 
consider a shift of the lattice coordinates 
$x\rightarrow x+\ell$, where $\ell$ is an arbitrary 
integer vector. It is easy to verify that the link 
variables $U^{\rm s}_\mu(x)$ and $U^{\rm s}_\mu(x+\ell)$ 
are related by 
\begin{eqnarray}
  \label{eq:lvtrl}
  U^{\rm s}_\mu(x+\ell)=\Omega^{\rm s}_\ell(x)
  U^{\rm s}_\mu(x)\Omega^{{\rm s}\dagger}_\ell(x+\hat\mu),
\end{eqnarray}
where $\Omega_\ell^{\rm s}(x)$ is given by
\begin{eqnarray}
  \label{eq:lsl}
  \Omega^{\rm s}_\ell(x)
  &=&\exp\left[-\frac{i}{2}
    \sum_{\mu,\nu}F_{\mu\nu}\ell_\mu x_\nu\right].
\end{eqnarray}
The relation (\ref{eq:lvtrl}) is analogous to a gauge
transformation. In general we cannot regard 
$U^{\rm s}_\mu(x+\ell)$ as a link variable associated 
with the link $(x,\hat\mu)$ since it does not 
satisfy the boundary conditions (\ref{eq:tbcsg}). 
However, eq. (\ref{eq:lvtrl}) suggests a transformation 
$T_\ell:\psi^{\rm s}\rightarrow\psi_\ell^{\rm s}$ defined by 
\begin{eqnarray}
  \label{eq:trpsil}
  T_\ell&:&
  \psi^{\rm s}_\ell(x)=\Omega^{{\rm s}\dagger}_\ell(x)
  \psi^{\rm s}(x+\ell). 
\end{eqnarray}
If $\psi^{\rm s}(x)$ is an eigenvector of the 
Hamiltonian and $\psi_\ell^{\rm s}(x)$ satisfies 
the twisted boundary conditions (\ref{eq:tbcsg}), 
then $\psi_\ell^{\rm s}(x)$ is also an eigenvector 
belonging to the same eigenvalue with $\psi^{\rm s}(x)$. 
In other words the transformation (\ref{eq:trpsil}) 
is a symmetry of the Hamiltonian. We call this 
as magnetic translation symmetry on the lattice. 

The requirement that (\ref{eq:trpsil}) satisfies 
the boundary conditions (\ref{eq:tbcsg}) leads 
to the consistency conditions for $\Omega_\ell^{\rm s}(x)$
\begin{eqnarray}
  \label{eq:ccl}
  \Lambda^{\rm s}_\mu(x+\ell)\Omega^{\rm s}_\ell(x)
  =\Omega^{\rm s}_\ell(x+L\hat\mu)\Lambda^{\rm s}_\mu(x).
\end{eqnarray}
This can be seen by noting that a translation 
$x\rightarrow x+L\hat\mu+\ell$ can be achieved 
in two different ways $x\rightarrow x+L\hat\mu
\rightarrow x+L\hat\mu+\ell$ and 
$x\rightarrow x+\ell\rightarrow x+\ell+L\hat\mu$. 
The conditions (\ref{eq:ccl}) can be stated in 
terms of $m_{\mu\nu}$ as 
\begin{eqnarray}
  \label{eq:mmnl}
  \sum_\nu m_{\mu\nu}\ell_\nu\equiv0\mod L.
\end{eqnarray}
One can also arrive at the same result by working in 
other gauges. In the axial gauge (\ref{eq:nplv})
magnetic translations are implemented by 
\begin{eqnarray}
  \label{eq:lal}
  \Omega^{\rm a}_\ell(x)&=&
  \exp\left[-i\sum_{\mu<\nu}F_{\mu\nu}\ell_\mu x_\nu\right].
\end{eqnarray}
We will use this later. 

We now turn to analyzing the constraints (\ref{eq:mmnl}). 
We can block-diagonalize $m_{\mu\nu}$  
into $2\times2$ anti-symmetric integer matrices
as in the continuum \cite{Igusa,Sakamoto-Tanimura,tif} 
\begin{eqnarray}
  \label{eq:m2LnLT}
  m_{\mu\nu}=\sum_{\rho,\sigma}\mathcal{L}_\mu{}^\rho \mathcal{L}_\nu{}^\sigma \nu_{\rho\sigma},
\end{eqnarray}
where $\mathcal{L}=(\mathcal{L}_\mu{}^\nu)$ is an integer matrix with $\det \mathcal{L}=1$ and 
$\nu=(\nu_{\mu\nu})$ takes the form
\begin{eqnarray}
  \label{eq:nmn}
  \nu=
  \begin{pmatrix}
    0 & \nu_1  &&&&& \\
    -\nu_1 & 0 &&&&& \\
    &&\ddots &&&&&\\
    &&&0&\nu_m&&& \\
    &&&-\nu_m&0&&& \\
    &&&&&0& \\
    &&&&&&\ddots&\\
    &&&&&&& 0
  \end{pmatrix}.
\end{eqnarray}
The set of integers $\nu_p$ ($p=1,\cdots,m$) with 
$2m$ being the rank of $(m_{\mu\nu})$ can be chosen so 
that $\nu_p$ divides $\nu_{p+1}$ for $p=1,\cdots,m-1$. 

We write $\mathcal{L}^{-1}$ in terms of $d$ integer 
vectors $M_p$, $N_p$ and $K_l$
($a=1,\cdots,m,~l=2m+1,\cdots,d$) as 
\begin{eqnarray}
  \label{eq:Linv}
  \mathcal{L}^{-1}=(M_1,N_1,\cdots,M_m,N_m,K_{2m+1},\cdots,K_d)^T
\end{eqnarray}
We also introduce the dual integer vectors  
$M^\ast{}^p$, $N^\ast{}^p$ and $K^\ast{}^l$ by
\begin{eqnarray}
  \label{eq:L}
  \mathcal{L}=(M^\ast{}^1,N^\ast{}^1,\cdots,M^\ast{}^m,N^\ast{}^m,
  K^\ast{}^{2m+1},\cdots,K^\ast{}^d). 
\end{eqnarray}
Then (\ref{eq:m2LnLT}) can be written as 
\begin{eqnarray}
  \label{eq:mnm2na}
  m_{\mu\nu}&=&\sum_{p=1}^m\nu_p(M_\mu^{\ast p}N_\nu^{\ast p}
  -N_\mu^{\ast p} M_\nu^{\ast p})
\end{eqnarray}
The set of $d$ integer vectors $M$, $N$ and $K$ also 
generates the original lattice. We can expand  
an arbitrary integer vector $\ell$ as
\begin{eqnarray}
  \label{eq:lMNK}
  \ell=\sum_{p=1}^m(\ell_M^pM_p+\ell_N^pN_p)
  +\sum_{l=1}^{d-2m}\ell_K^lK_l, 
\end{eqnarray}
where $\ell_M^p$, $\ell_N^p$ and $\ell_K^l$ are all integers. 
The constraints (\ref{eq:mmnl}) for the magnetic translations $\ell$ 
can be simplified as
\begin{eqnarray}
  \label{eq:lscod}
  \nu_p \ell_M^p\equiv\nu_p\ell_N^p\equiv0 \mod L, 
\end{eqnarray}
whereas $\ell_K^l$ can be arbitrary. 

To solve the conditions (\ref{eq:lscod}) 
we introduce a set of integers $r_p$, $s^p>0$ and $n^p$ by 
\begin{eqnarray}
  \label{eq:nprpsp}
  n^pr_p=\nu_p, \qquad r_ps^p=L
\end{eqnarray}
where $n^p$ and $s^p$ are mutually prime. Then
$\ell_M^p=js^p$ and $\ell_N^p=ks^p$ satisfy (\ref{eq:lscod}) 
for arbitrary integers $j$, $k$. We thus find three 
basic vectors generating magnetic translations
\begin{eqnarray}
  \label{eq:tbv}
  \ell^{(M)}_p=s^pM_p, \quad
  \ell^{(N)}_p=s^pN_p, \quad
  \ell^{(K)}_l=K_l.
\end{eqnarray}
In the following sections we need another set of integers 
$r'_p$, $s'_p>0$ and $n^{\prime p}$ defined by 
\begin{eqnarray}
  \label{eq:npprppspp}
  n^{\prime p}s'_p=n^p, \qquad 
  r'_ps'_p=r_p, 
\end{eqnarray}
where $n^{\prime p}$ and $r'_p$ are mutually prime.  

\section{Magnetic translations on 2d lattice}
\label{sec:mtr2dl}
\setcounter{equation}{0}

To illustrate magnetic translations on the lattice 
we consider a two dimensional system with a magnetic 
field $F_{12}=2\pi\nu/L^2$. This is already of a standard 
form (\ref{eq:m2LnLT}) and $\mathcal{L}$ is taken to be the 
identity matrix. The integer vectors $M_1$, $N_1$ and 
their dual are given by 
\begin{eqnarray*}
  M_1=M^{\ast1}=
  \begin{pmatrix}
    1 \\ 0
  \end{pmatrix},
  \qquad
  N_1=N^{\ast1}=
  \begin{pmatrix}
    0 \\ 1
  \end{pmatrix}. 
\end{eqnarray*}

As in (\ref{eq:nprpsp}),  we have a unique set of 
integers $n=n^1$, $r=r_1$ and $s=s^1$ for a given 
$L$ and $\nu=\nu_1$.  
We also define another set of integers 
$n'=n^{\prime1}$, $r'=r'_1$  
and $s'=s^{\prime1}$ by (\ref{eq:npprppspp}). 
We thus find two independent basic magnetic 
translations $T_x\equiv T_{sM_1}$ and $T_y\equiv T_{sN_1}$ corresponding 
to $x\rightarrow x+s M_1$ and $x\rightarrow x+s N_1$, 
respectively. In the axial gauge these act on the
wave functions by
\begin{eqnarray}
  \label{eq:mtropsi}
  T_x\psi(x,y)=e^{\frac{2i\pi n'y}{sr'}}\psi(x+s,y), 
  \qquad T_y\psi(x,y)=\psi(x,y+s),
\end{eqnarray}
where use has been made of (\ref{eq:lal}). We have 
suppressed the label ``a'' standing for the axial 
gauge. As can be verified from (\ref{eq:mtropsi}), 
these operator satisfy 
\begin{eqnarray}
  \label{eq:txty}
  T_xT_y=e^{-\frac{2i\pi n'}{r'}}T_yT_x. 
\end{eqnarray}
This gives $T_x^{r'}T_y=T_yT_x^{r'}$. We can find 
simultaneous eigenvectors of $H$, $T_x^{r'}$ 
and $T_y$. The eigenvalues of $T_x^{r'}$ can be 
written as $e^{\frac{2i\pi q}{s'}}$ by an integer 
$q$ ($0\leq q<s'$) since $T_x$ satisfies 
$(T_x^{r'})^{s'}=T_x^r=1$. Similarly, we can 
write the eigenvalue of $T_y$ as $e^{\frac{2i\pi p}{r}}$ 
with an integer $0\leq p<r$. Let us denote a 
simultaneous eigenvectors by $\psi_{p,q}$, then 
we have 
\begin{eqnarray}
  \label{eq:evs}
  &&H\psi_{p,q}(x,y)=\lambda_{p,q}
  \psi_{p,q}(x,y), \nonumber \\
  &&T_x^{r'}\psi_{p,q}(x,y)=e^{\frac{2i\pi q}{s'}}\psi_{p,q}(x,y), \\
  &&T_y\psi_{p,q}(x,y)=e^{\frac{2i\pi p}{r}}\psi_{p,q}(x,y).
  \nonumber 
\end{eqnarray}
The magnetic translation $T^m_x$ ($m=0,1,\cdots,r'-1$) 
maps an eigenstate of $T_y$ belonging to an eigenvalue 
$e^{\frac{2i\pi p}{r}}$ to other eigenstate with an 
eigenvalue $e^{\frac{2i\pi(p+mn)}{r}}$. In other words, 
it changes the label $p\mod r$ to $p+mn\mod r$. Since 
$p+r'n\equiv p\mod r$, there are $r'$  degenerate 
eigenvectors belonging to the eigenvalue 
$\lambda_{p,q}$. We thus find that the eigenvalue 
$\lambda_{p,q}$ depends only on $p\mod s'$ with 
respect to $p$. 

The second and the third of (\ref{eq:evs}) together 
with (\ref{eq:mtropsi}) imply that the wave 
functions at $(x,y)$ and $(x+sr',y)$ or $(x,y+s)$ 
are related by 
\begin{eqnarray}
  \label{eq:trpsiprop}
  \psi_{p,q}(x+sr',y)=e^{-\frac{2i\pi n'y}{s}+\frac{2i\pi q}{s'}}
  \psi_{p,q}(x,y), \qquad
  \psi_{p,q}(x,y+s)=e^{\frac{2i\pi p}{r}}\psi_{p,q}(x,y). 
\end{eqnarray}
Noting the twisted periodicity in $y$, we can 
expand $\psi_{p,q}(x,y)$ in Fourier series as 
\begin{eqnarray}
  \label{eq:ps2ph}
  \psi_{p,q}(x,y)&=&\frac{1}{\sqrt{s}}\sum_{j=0}^{s-1}
  \varphi_{p,q;j}(x)e^{\frac{2i\pi(p+jn'r)y}{L}}, 
\end{eqnarray}
where $\varphi_{p,q;j}(x)$ must satisfy
\begin{eqnarray}
  \label{eq:vpprop}
  \varphi_{p,q;j+s}(x)=\varphi_{p,q;j}(x), \qquad
  \varphi_{p,q;j}(x+sr')=e^{\frac{2i\pi q}{s'}}\varphi_{p,q;j+1}(x). 
\end{eqnarray}
We see that the original $L^2$ components of 
$\psi_{p,q}(x,y)$ ($0\leq x,y<L$) can be expressed 
in terms $s^2r'$ values of a one dimensional wave 
function $\varphi_{p,q}(x)\equiv\varphi_{p,q;0}(x)$ 
($0\leq x<s^2r'$). It satisfies the twisted 
periodicity 
\begin{eqnarray}
  \label{eq:twvpkl}
  \varphi_{p,q}(x+s^2r')=e^{\frac{2i\pi sq}{s'}}
  \varphi_{p,q}(x) 
\end{eqnarray}
as one can see from (\ref{eq:vpprop}). The magnetic 
translations put no further restrictions on 
$\varphi_{p,q}(x)$. 

Our arguments so far does not depend on the 
detailed form of the Hamiltonian $H$. We can 
block-diagonalize it into $r'$ matrices 
whatever form it is. This leads us to the 
Hamiltonian $\mathcal{H}_{p;j}$ defined by 
\begin{eqnarray}
  \label{eq:redD}
  H\psi_{p,q}(x,y)
  &=&\frac{1}{\sqrt{s}}\sum_{j=0}^{s-1}e^{\frac{2i\pi(p+jn'r)y}{L}}
  {\cal H}_{p;j}\varphi_{p,q;j}(x). 
\end{eqnarray} 
It acts on the one dimensional system of a size 
$s^2r'$ satisfying the twisted 
boundary condition (\ref{eq:twvpkl}). Since the 
eigenvalues of $\mathcal{H}_{p;j}$ do not 
depend on $j$, it is enough to consider the 
case $j=0$ as mentioned above. We thus arrive 
at the eigenvalue equation for $\varphi_{p,q}(x)$ 
\begin{eqnarray}
  \label{eq:eveqvp}
  \mathcal{H}_p\varphi_{p,q}(x) 
  =\lambda_{p,q}\varphi_{p,q}(x), 
\end{eqnarray}
where $\mathcal{H}_p$ stands for 
$\mathcal{H}_{p;0}$. 

\begin{figure}[t]
  \centering
  \epsfig{file=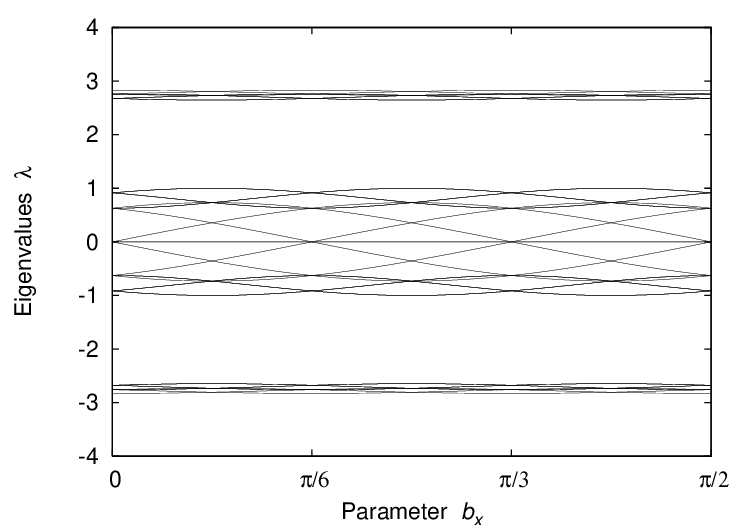,clip=,height=8cm,angle=0}
  \caption{Eigenvalues of tight-binding Hamiltonian 
    in two dimensions are plotted for $0\leq b_x\leq 
    \pi/2$ and $b_y=0$. We take $L=6$, $\nu=9$. 
  }
  \label{fig:bx}
\end{figure}

For tight-binding  Hamiltonian $\psi_{p,q}(x,y)$ 
satisfies  
\begin{eqnarray}
  \label{eq:tbh}
  &&e^{ib_x}\psi_{p,q}(x+1,y)+e^{-ib_x}\psi_{p,q}(x-1,y)
  \nonumber \\
  && \hskip 1cm +e^{\frac{2i\pi n' x}{s^2r'}+ib_y}\psi_{p,q}(x,y+1)
  +e^{-\frac{2i\pi n'x}{s^2r'}-ib_y}\psi_{p,q}(x,y-1)
  =\lambda_{p,q}\psi_{p,q}(x,y). 
\end{eqnarray}
The equations for $\varphi_{p,q}(x)$ can be 
found by inserting (\ref{eq:ps2ph}) into the 
expression. The constant phase factors $e^{\pm ib_x}$ 
can be eliminated if we introduce 
\begin{eqnarray}
  \label{eq:tivp}
  \tilde\varphi_{p,q}(x)
  =e^{ib_xx}\varphi_{p,q}(x). 
\end{eqnarray}
We thus obtain   
\begin{eqnarray}
  \label{eq:rheq}
  \tilde\varphi_{p,q}(x+1)+\tilde\varphi_{p,q}(x-1)
  +\left\{2\cos\left(\frac{2\pi n'x}{s^2r'}
      +\frac{2\pi p}{sr}+b_y\right)-\lambda_{p,q}
    \right\}\tilde\varphi_{p,q}(x)=0. 
\end{eqnarray}
This is referred to as the Harper equation \cite{Harp} 
in condensed matter physics. The periodicity (\ref{eq:twvpkl}) 
under the shift $x\rightarrow x+s^2r'$ is further 
twisted by (\ref{eq:tivp}) and is given by 
\begin{eqnarray}
  \label{eq:tbctvp}
  \tilde\varphi_{p,q}(x+s^2r')=e^{\frac{2i\pi sq}{s'}+ib_xs^2r'}
  \tilde\varphi_{p,q}(x). 
\end{eqnarray}
The plot of the spectrum as a function of 
$\alpha
=n'/s^2r'$, i.e., magnetic flux per plaquette,  
is known as the butterfly 
diagram \cite{hof}. 
Unlike the continuum theories eigenvalues depends on 
the parameters $b_x$ and $b_y$. The twisted periodicity 
(\ref{eq:tbctvp}) is invariant under the shift 
$b_x\rightarrow b_x+2\pi/s^2r'$. This implies that 
the eigenvalues are periodic in $b_x$ and 
also in $b_y$ by rotation symmetry with a 
period $2\pi/s^2r'$. See Fig.\ref{fig:bx}. 

\section{Extension to higher dimensions}
\label{sec:olc}
\setcounter{equation}{0}

We have shown that the magnetic translation 
symmetry can be used to constrain the 
wave functions on a two dimensional square lattice. 
We now extend this to higher dimensions. 

Magnetic translation symmetry is apparent for special 
types of magnetic fields like (\ref{eq:nmn}) and 
the analysis for the two dimensional system in the 
previous section is applicable immediately. 
This is not the case for a general uniform magnetic 
field. Fortunately, it is always possible to transform 
$F_{\mu\nu}$ into the block-diagonal form (\ref{eq:nmn}) 
as mentioned in Sect. \ref{sec:lmtr}. In the 
continuum eigenvalue problems on magnetized tori 
can be solved by using this approach \cite{Baal}. 
In this section we use the notation of Ref. \cite{tif}.

We introduce oblique lattice coordinates 
$\xi^p,~\eta^p,~\chi^l$ ($p=1,\cdots,
m$, $l=2m+1,\cdots,d$) by 
\begin{eqnarray}
  \label{eq:olc}
  x=\sum_p(M_p\xi^p+N_p\eta^p)+\sum_lK_l\chi^l,
\end{eqnarray}
where $M_p$, $N_p$ and 
$K_l$ are oblique lattice vectors defined by 
(\ref{eq:Linv}). 
The oblique lattice coordinates are all integers 
and have one-to-one correspondence with $x$ as 
can be seen from 
\begin{eqnarray}
  \label{eq:xhc2x}
  \xi^p=\sum_\mu M^{\ast p}_\mu x_\mu, \qquad
  \eta^p=\sum_\mu N_\mu^{\ast p}x_\mu, \qquad
  \chi^l=\sum_\mu K^{\ast l}_\mu x_\mu.
\end{eqnarray}
In particular the unit translations 
$x\rightarrow x\pm\hat\mu$ correspond to the 
following shifts 
\begin{eqnarray}
  \label{eq:utrxxhc}
  \xi^p \rightarrow \xi^p\pm M^{\ast p}_\mu, \qquad
  \eta^p \rightarrow \eta^p\pm N^{\ast p}_\mu, \qquad
  \chi^l \rightarrow \chi^l\pm K^{\ast l}_\mu.
\end{eqnarray} 
Conversely, the translation $\xi,~\eta,~\chi
\rightarrow\xi\pm e_p,~\eta,~\chi$ can be realized 
by the shift $x \rightarrow x\pm M_p$, where 
$e_p$ is the $p$-th unit vector with $(\xi+e_p)_q
=\xi_q+\delta_{p,q}$. Similar things hold true 
for the unit translations in $\eta$ or in $\chi$. 

The link variables in the symmetric gauge 
(\ref{eq:lvsg}) can be written as 
\begin{eqnarray}
  \label{eq:lvsolc}
  U^{{\rm s}}_{\mu}(x)&=&\exp\left[
    -\frac{i\pi}{L^2}\sum_p\nu_p(M_\mu^{\ast p}\eta^p
    -N_\mu^{\ast p}\xi^p)+ib_\mu\right] \nonumber \\
  &=&\exp\left[
    -i\sum_p\frac{\pi n^{\prime p}}{(s^p)^2r'_p}(M_\mu^{\ast p}\eta^p
    -N_\mu^{\ast p}\xi^p)+ib_\mu\right], 
\end{eqnarray}
where $n^{\prime p}$, $r'_p$ and $s^p$ are given 
by (\ref{eq:nprpsp}) and (\ref{eq:npprppspp}).
To argue magnetic translations it is more 
convenient to work in axial gauge on the oblique 
lattice. This can be achieved by the following 
gauge transformation 
\begin{eqnarray}
  \label{eq:b2ag}
  && U^{\alpha}_{\mu}(x)=\Lambda^{\alpha{\rm s}}(x)
  U^{{\rm s}}_\mu(x)\Lambda^{\alpha{\rm s}\dagger}(x+\hat\mu)
  =\exp\left[i\sum_p\frac{2\pi n^{\prime p}}{(s^p)^2r'_p}
    N^{\ast p}_\mu\left(\xi^p+\frac{1}{2}
    M^{\ast p}_\mu\right)+ib_\mu\right] \nonumber\\
  &&\hbox{with} \quad\Lambda^{\alpha{\rm s}}(x)
  =\exp\left[-i
    \sum_p\frac{\pi n^{\prime p}}{(s^p)^2r'_p}\xi^p\eta^p\right].
\end{eqnarray}
In this gauge the twisted boundary conditions for the 
wave functions are given by 
\begin{eqnarray}
  \label{eq:paLperi}
  && \psi(x+L\hat\mu)
  =\Lambda^{\alpha}_\mu(x)\psi(x) \nonumber \\
  &&\hbox{with}\quad \Lambda^{\alpha}_\mu(x)
  =\exp\left[-i\sum_p\frac{2\pi n^p}{s^p}
    M_\mu^{\ast p}
    \left(\eta^p+\frac{L}{2}N_\mu^{\ast p}\right)\right].
\end{eqnarray}
We have suppressed the label $\alpha$ indicating the 
axial gauge for the wave function. 

We can also find the periodicity of the wave function 
under shifts of the oblique lattice coordinates by $L$ 
by a repeated use of (\ref{eq:paLperi}) as
\begin{eqnarray}
  \label{eq:pxitr}
  \psi(\xi+Le_p,\eta,\chi)&=&e^{-\frac{2i\pi n^p}{s^p}
    \eta^p+i\pi\epsilon^{(M)}_p}\psi(\xi,\eta,\chi), 
  \nonumber\\
  \psi(\xi,\eta+Le_p,\chi)&=&e^{i\pi\epsilon^{(N)}_p}
  \psi(\xi,\eta,\xi), \\
  \psi(\xi,\eta,\chi+Le_l)&=&e^{i\pi\epsilon^{(K)}_l}
  \psi(\xi,\eta,\chi). \nonumber
\end{eqnarray}
where $\epsilon$'s are integers defined by
\begin{eqnarray}
  \label{eq:eKl}
  \epsilon^{(M)}_p=\sum_{\mu<\nu}m_{\mu\nu}M_{p\mu}M_{p\nu}, 
  \quad 
  \epsilon^{(N)}_p=\sum_{\mu<\nu}m_{\mu\nu}N_{p\mu}N_{p\nu}, 
  \quad
  \epsilon^{(K)}_l=\sum_{\mu<\nu}m_{\mu\nu}K_{l\mu}K_{l\nu}. 
\end{eqnarray}
We see that the wave function is either periodic or anti-periodic 
in $\eta^p$ and $\chi^l$. 

We now turn to magnetic translation of the wave function.
In the axial gauge (\ref{eq:b2ag}) the link 
variables on the sites $x$ and $x+\ell$ are related 
by  
\begin{eqnarray*}
  && U^\alpha_\mu(x+\ell)=\Omega^\alpha_\ell(x)
  U_\mu^\alpha(x)\Omega_\ell^{\alpha\dagger}(x+\hat\mu) \nonumber \\
  &&\hbox{with}\quad 
  \Omega_\ell^\alpha(x)=\exp\left[-i
    \sum_p\frac{2\pi n^{\prime p}}{(s^p)^2r_p'}\ell_M^p\eta^p\right], 
\end{eqnarray*}
where $\ell$ is given by (\ref{eq:lMNK}). This leads to 
magnetic translation of the wave function
\begin{eqnarray}
  \label{eq:magtroag}
  T_\ell\psi(\xi,\eta,\chi)
  =\Omega_\ell^{\alpha\dagger}(x)\psi(\xi+\ell_M,\eta+\ell_N,\chi+\ell_K). 
\end{eqnarray}
Let us denote the generators of magnetic 
translations corresponding to the three basic 
translations (\ref{eq:tbv}) by $T_p^{(M)}$, 
$T_p^{(N)}$ and $T_l^{(K)}$, respectively. 
Then (\ref{eq:magtroag}) yields 
\begin{eqnarray}
  \label{eq:bmagtroag}
  && T_p^{(M)}\psi(\xi,\eta,\chi)
  =\exp\left[\frac{2i\pi n^{\prime p}}{s^pr_p'}\eta^p\right]
  \psi(\xi+s^pe_p,\eta,\chi), \nonumber \\
  &&T_p^{(N)}\psi(\xi,\eta,\chi)=\psi(\xi,\eta+s^pe_p,\chi), \\
  &&T_l^{(K)}\psi(\xi,\eta,\chi)=\psi(\xi,\eta,\chi+e_l), \nonumber
\end{eqnarray}
The generators of magnetic translations satisfy 
\begin{eqnarray}
  \label{eq:commTs}
  T^{(M)}_pT_p^{(N)}=e^{-\frac{2i\pi n^{\prime p}}{r_p'}}
  T_p^{(N)}T_p^{(M)}
\end{eqnarray}
and all other combinations are commutative. They also satisfy 
\begin{eqnarray}
  \label{eq:tneq1}
  ((T_p^{(M)})^{r_p'})^{s'_p}=e^{i\pi\epsilon^{(M)}_p}, 
  \quad (T^{(N)}_p)^{r_p}=e^{i\pi\epsilon^{(N)}_p}, \quad
  (T^{(K)}_l)^L=e^{i\pi\epsilon^{(K)}_l}, 
\end{eqnarray}
as can be seen from (\ref{eq:pxitr}) and (\ref{eq:bmagtroag}). 
These are a higher dimensional generalization of (\ref{eq:mtropsi})
and (\ref{eq:txty}). 

We may choose a set of commuting operators $H$, 
$(T_p^{(M)})^{r_p'}$, $T^{(N)}_p$, $T^{(K)}_l$ 
as in two dimensions and consider eigenvectors 
of these operators 
\begin{eqnarray}
  \label{eq:Tsreal}
  &&H\psi_{k,p,q}(\xi,\eta,\chi)
  =\lambda_{k,p,q}\psi_{k,p,q}(\xi,\eta,\chi), \nonumber\\
  &&(T_p^{(M)})^{r_p'}\psi_{k,p,q}(\xi,\eta,\chi)
  =\exp\left[\frac{2i\pi}{s'_p}
    \left(q_p+\frac{1}{2}\epsilon_p^{(M)}\right)\right]
    \psi_{k,p,q}(\xi,\eta,\chi), \nonumber\\
  &&T_p^{(N)}\psi_{k,p,q}(\xi,\eta,\chi)
  =\exp\left[\frac{2i\pi}{r_p}
    \left(p_p+\frac{1}{2}\epsilon_p^{(N)}\right)\right]
  \psi_{k,p,q}(\xi,\eta,\chi), \nonumber\\
  &&T^{(K)}_l\psi_{k,p,q}(\xi,\eta,\chi)
  =\exp\left[\frac{2i\pi}{L}
    \left(k_l+\frac{1}{2}\epsilon^{(K)}_l\right)\right]
  \psi_{k,p,q}(\xi,\eta,\chi),
\end{eqnarray}
where $0\leq q_p<s_p'$, $0\leq p_p<r_p$ and $0\leq k_l<L$ 
are integers. These together with (\ref{eq:bmagtroag}) 
implies
\begin{eqnarray}
  \label{eq:tperisrp}
  &&\psi_{k,p,q}(\xi+s^pr'_pe_p,\eta,\chi)
  =\exp\left[-\frac{2i\pi n^{\prime p}}{s^p}\eta^p
    +\frac{2i\pi}{s_p'}\left(q_p+\frac{1}{2}\epsilon_p^{(M)}
    \right)\right]\psi_{k,p,q}(\xi,\eta,\chi), \nonumber \\
  &&\psi_{k,p,q}(\xi,\eta+s^pe_p,\chi)
  =\exp\left[\frac{2i\pi}{r_p}
    \left(p_p+\frac{1}{2}\epsilon_p^{(N)}\right)\right]
  \psi_{k,p,q}(\xi,\eta,\chi), \\
  &&\psi_{k,p,q}(\xi,\eta,\chi+e_l)=\exp\left[\frac{2i\pi}{L}
    \left(k_l+\frac{1}{2}\epsilon^{(K)}_l\right)\right]
  \psi_{k,p,q}(\xi,\eta,\chi). \nonumber
\end{eqnarray}
The last two of these expressions enable us to write  
\begin{eqnarray}
  \label{eq:ps2ps1}
  \psi_{k,p,q}(\xi,\eta,\chi)
    =\frac{1}{\sqrt{s^1\cdots s^m}}\sum_j\varphi_{k,p,q;j}(\xi)
    e^{\frac{2i\pi}{L}\left\{\sum_q\left(n^{\prime q}r_qj_q+p_q
        +\frac{1}{2}\epsilon^{(N)}_q\right)\eta^q
      +\sum_l\left(k_l+\frac{1}{2}\epsilon^{(K)}_l\right)\chi^l\right\}},
\end{eqnarray}
where the sum is taken over integers $0\leq j_p<s^p$ 
($p=1,\cdots,m$). The Fourier coefficients 
$\varphi_{k,p,q;j}(\xi)$ must satisfy the following periodicity 
\begin{eqnarray}
  \label{eq:vpkpqperi}
  \varphi_{k,p,q;j+s^pe_p}(\xi)=\varphi_{k,p,q;j}(\xi), \qquad
  \varphi_{k,p,q;j}(\xi+s^pr'_pe_p)
  =e^{\frac{2i\pi}{s'_p}\left(q_p+\frac{1}{2}\epsilon^{(M)}_p\right)}
  \varphi_{k,p,q;j+e_p}(\xi)
\end{eqnarray}
The first comes from the definition of $\varphi_{k,p,q;j}(\xi)$ 
and the second from the first of (\ref{eq:tperisrp}). 
We see that the original $L^d$ components of the wave 
function $\psi_{k,p,q}(x)$ ($0\leq x_\mu<L$, $\mu=1,\cdots,d$) 
can be obtained from the $(s^1)^2r'_1\cdots(s^m)^2r'_m$ components of 
$\varphi_{k,p,q}(\xi)\equiv\varphi_{k,p,q;0}(\xi)$ 
($0\leq \xi^p<(s^p)^2r'_p$, $p=1,\cdots,m$). It must satisfy 
the twisted boundary conditions
\begin{eqnarray}
  \label{eq:tpvpmdim}
  \varphi_{k,p,q}(\xi+(s^p)^2r'_pe_p)
  =e^{\frac{2i\pi s^p}{s'_p}\left(q_p+\frac{1}{2}\epsilon^{(M)}_p\right)}
  \varphi_{k,p,q}(\xi)
\end{eqnarray}
and can be considered as a wave functions of some $m$ 
dimensional system. The Hamiltonian ${\cal H}_{k,p;j}$ 
of the dimensionally reduced system is found by inserting 
(\ref{eq:ps2ps1}) into the first of (\ref{eq:Tsreal}), i.e., 
\begin{eqnarray}
  \label{eq:redsysham}
  H\psi_{k,p,q}(\xi,\eta,\chi)
  =\frac{1}{\sqrt{s^1\cdots s^m}}\sum_j
    e^{\frac{2i\pi}{L}\left\{\sum_q\left(n^{\prime q}r_qj_q+p_q
        +\frac{1}{2}\epsilon^{(N)}_q\right)\eta^q
      +\sum_l\left(k_l+\frac{1}{2}\epsilon^{(K)}_l\right)\chi^l\right\}}
  {\cal H}_{k,p;j}\varphi_{k,p,q;j}(\xi). \nonumber \\
\end{eqnarray}
It is enough to consider the case $j=0$. We thus obtain 
\begin{eqnarray}
  \label{eq:eveqCHvp}
  {\cal H}_{k,p}\varphi_{k,p,q}(\xi)=\lambda_{k,p,q}
  \varphi_{k,p,q}(\xi),
\end{eqnarray}
where we have defined ${\cal H}_{k,p}
\equiv{\cal H}_{k,p;0}$. This together with 
(\ref{eq:tpvpmdim}) determines the wave functions 
$\varphi_{k,p,q}$, or equivalently 
$\varphi_{k,p,q;j}$, and the eigenvalues 
$\lambda_{k,p,q}$. They are defined on 
the $m$ dimensional oblique lattice of a size 
$(s^1)^2r'_1\times\cdots\times
(s^m)^2r_m'$. We can then reconstruct 
$\psi_{k,p,q}$ by (\ref{eq:ps2ps1}) and arrive at 
$r_1\cdots r_ms_1'\cdots s_m'L^{d-2m}$ 
wave functions forming a complete set. 

For any fixed $k$ and $q$ there are $r_1\cdots r_m$ 
wave functions $\psi_{k,p,q}$. They can be decomposed 
into $s_1'\cdots s_m'$ sets of wave 
functions degenerated by the magnetic translation 
symmetry. For a given set of labels $p_q\mod s'_q$ 
($q=1,\cdots,m$) there are exactly $r_1'\cdots r'_m$ 
degenerate wave functions given by 
\begin{eqnarray}
  \label{eq:degwfbmt}
  (T_1^{(M)})^{h_1}\cdots(T_m^{(M)})^{h_m}\psi_{k,p,q}(\xi). 
  \qquad(0\leq h_q<r'_q,~q=1,\cdots,m)
\end{eqnarray}

We now apply the formalism developed so far to 
the tight-binding system described by 
\begin{eqnarray}
  \label{eq:ddimtbs}
  \sum_{\mu=1}^d\{U^\alpha_\mu(x)\psi_{k,p,q}(x+\hat\mu)
  +U_\mu^{\alpha\dagger}(x-\hat\mu)\psi_{k,p,q}(x-\hat\mu)\}
  =\lambda_{k,p,q}\psi_{k,p,q}(x), 
\end{eqnarray}
where the link variables are given by (\ref{eq:b2ag}). 
This can be converted to the form (\ref{eq:eveqCHvp}). 
To have a compact expression we introduce $\beta$'s 
by the expansion 
\begin{eqnarray}
  \label{eq:beta}
  b_\mu=\sum_q(\beta^{(M)}_qM^{\ast q}_\mu
  +\beta^{(N)}_qN^{\ast q}_\mu)+\sum_l\beta^{(K)}_lK^{\ast l}_\mu, 
\end{eqnarray}
and define $\tilde p$, $\tilde q$ and $\tilde k$ by
\begin{eqnarray}
  \label{eq:tptqtk}
  \tilde p_p=p_p+\frac{1}{2}\epsilon^{(N)}_p
  +\frac{L}{2\pi}\beta^{(N)}_p, \quad
  \tilde q_p=q_p+\frac{1}{2}\epsilon^{(M)}_p
  +\frac{L}{2\pi}\beta^{(M)}_p, \quad
  \tilde k_l=k_l+\frac{1}{2}\epsilon^{(K)}_l
      +\frac{L}{2\pi}\beta^{(K)}_l.
\end{eqnarray}
Then (\ref{eq:ddimtbs}) is reduced to 
\begin{eqnarray}
  \label{eq:gheq}
  &&\sum_{\mu=1}^d\bigl\{
  e^{\frac{2i\pi}{L}\sum_qN_\mu^{\ast q}
    \left\{\frac{\nu_q}{L}
      \left(\xi^q+\frac{1}{2}M_\mu^{\ast q}\right)
      +\tilde p_q\right\}
    +\frac{2i\pi}{L}
    \sum_l K_\mu^{\ast l}\tilde k_l+i\sum_q\beta^{(M)}_qM^{\ast q}_\mu}
  \varphi_{k,p,q}(\xi+M_\mu^\ast) \nonumber \\
  &&\hskip .5cm+e^{-\frac{2i\pi}{L}\sum_qN_\mu^{\ast q}
  \left\{\frac{\nu_q}{L}
  \left(\xi^q-\frac{1}{2}M_\mu^{\ast q}\right)
  +\tilde p_q\right\}
  -\frac{2i\pi}{L}
  \sum_l K_\mu^{\ast l}\tilde k_l
  -i\sum_q\beta^{(M)}_qM^{\ast q}_\mu}\varphi_{k,p,q}(\xi-M_\mu^\ast)
  \bigr\}=\lambda_{k,p,q}\varphi_{k,p,q}(\xi). \nonumber \\
\end{eqnarray}
The phase factor containing $\beta^{(M)}$ can be 
removed by introducing $\tilde\varphi_{k,p,q}$ as
\begin{eqnarray}
  \label{eq:tilvpkpq}
  \tilde\varphi_{k,p,q}(\xi)
  &=&e^{i\sum_q\beta^{(M)}_q\xi^q}\varphi_{k,p,q}(\xi).
\end{eqnarray}
We finally obtain 
\begin{eqnarray}
  \label{eq:gheqtvp}
  &&\sum_{\mu=1}^d\Biggl\{
  e^{\frac{2i\pi}{L}\sum_qN_\mu^{\ast q}
    \left\{\frac{\nu_q}{L}
      \left(\xi^q+\frac{1}{2}M_\mu^{\ast q}\right)
      +\tilde p_q\right\}
    +\frac{2i\pi}{L}
    \sum_l K_\mu^{\ast l}\tilde k_l}
  \tilde\varphi_{k,p,q}(\xi+M_\mu^\ast) \nonumber \\
  &&\hskip 1cm+e^{-\frac{2i\pi}{L}\sum_qN_\mu^{\ast q}
    \left\{\frac{\nu_q}{L}
      \left(\xi^q-\frac{1}{2}M_\mu^{\ast q}\right)
      +\tilde p_q\right\}
    -\frac{2i\pi}{L}
    \sum_l K_\mu^{\ast l}\tilde k_l}
  \tilde\varphi_{k,p,q}(\xi-M_\mu^\ast)
  \Biggr\}=\lambda_{k,p,q}\tilde\varphi_{k,p,q}(\xi), 
  \nonumber \\
  &&\hbox{with}\quad
  \tilde\varphi_{k,p,q}(\xi+(s^p)^2r'_pe_p)
  =e^{\frac{2i\pi s^p}{s'_p}\tilde q_p}
  \tilde\varphi_{k,p,q}(\xi)
\end{eqnarray}
This is a higher dimensional extension  
of the Harper equation (\ref{eq:rheq}). 
The twisted periodicity can be derived from 
(\ref{eq:tpvpmdim}) and (\ref{eq:tilvpkpq}). 
It is periodic under the shift $\beta^{(M)}\rightarrow
\beta^{(M)}+2\pi e_p/(s^p)^2r'_p$. 
This implies that the spectrum is symmetric 
under the same shift as has been observed in 
two dimensions.

\section{Summary and discussion}
\label{sec:sd}

We have investigated magnetic translation 
symmetries on finite periodic lattices in arbitrary 
dimensions. We have shown that any lattice 
system with uniform background magnetic field 
possesses the symmetry. In continuum 
theories magnetic flux 
tensor $m_{\mu\nu}$ determines a unique magnetic 
translation group. In lattice theories, however, 
it depends on how the lattice size $L$ is 
taken. This is due to the simple fact that 
only translations by a divisor of $L$ is 
allowed on the lattice whereas $\nu_p$ is not 
necessary a divisor of $L$. The symmetry changes 
rapidly as one varies the magnetic field. Then 
the spectrum of $H$ is very sensitive with respect 
to the applied magnetic field. These qualitative 
differences between the continuum and the lattice 
disappear in the continuum limit. 
We can always take the limit keeping any prescribed 
magnetic translation symmetry. 

In the extreme case that there is no magnetic 
translation shorter than the period of the lattice, 
eigenvectors cannot be constrained by the magnetic 
translation symmetries. We cannot reduce the number 
of unknown variables in solving the eigenvalue 
problem. However, it is always possible to 
dimensionally reduce the system. An immediate 
consequence of this is that any tight-binding 
system can be formulated by the generalized Harper 
equations. Dimensionally reduced systems may be more 
tractable than the original equations and be 
beneficial to further studies. 

As an application of magnetic translation symmetry
we can compute index of overlap Dirac operator for 
abelian gauge background. It is related to 
spectral asymmetry of hermitian Wilson-Dirac 
operator. Due to the topological invariance of 
the index it suffices to compute the spectral 
asymmetry for a uniform magnetic field belonging 
to the same topological sector. Since the 
degeneracy of each eigenvector is easily 
seen from the magnetic translation symmetry, 
we have only to compute the spectral asymmetry 
of the reduced Hamiltonian. This has been 
carried out in two dimensions. Extension to 
higher dimensions is certainly interesting. We 
will argue the magnetic translation symmetry 
in lattice fermion systems elsewhere.

\end{document}